# Resurrecting the One-sided P-value as a Likelihood Ratio

Nicholas Adams, Alfred Health

**Abstract**

The one-sided P-value has a long history stretching at least as far back as Laplace (1812) but has in recent times been mostly supplanted by the two-sided P-value. We present justification for a bijective relationship between the one-sided P-value and a likelihood ratio based on maximum likelihood, a relationship that cannot be demonstrated for the two-sided P-value. A number of criticisms of P-values are discussed and it is shown that many of these criticisms are not justified when a likelihood ratio interpretation of a one-sided P-value is employed. Converting a one-sided P-value to a likelihood ratio provides the advantages of the likelihood evidential paradigm.

1. **INTRODUCTION**

It is perhaps paradoxical that the most widely used statistic, the P-value, is also the most widely criticised. We briefly consider some of these criticisms below, but it is worth noting here that most are directed at the two-sided P-value rather than the one-sided. For an isomorphic and symmetric distribution like the normal, the one-sided P-value is exactly one half of the two-sided and this leads to the misapprehension that the one-sided form is merely a weaker, or less conservative, form of the two-sided. However, we argue that they are fundamentally different, the two-sided P-value being a test of existence and the one-sided being, (when properly employed), a test of direction and magnitude. In the dominant null hypothesis significance testing framework, the point null is assumed to be true, and any discordance between the data and this hypothesis is implied as being due to the play of chance, and hence its direction is of no interest. Of course, the direction of the effect is obvious to us, but the point is, the strength of evidence supporting the effect being in that direction is not fully captured by the two-sided P-value. By contrast, in the framework of a one-sided P-value against a dividing hypothesis the assumption is of a non-zero effect in one direction or the other, and hence both the direction and magnitude of an observed effect are of interest. In this paper we derive a bijective relationship between the one-sided P-value and a likelihood ratio, a relationship that cannot be demonstrated for the two-sided P-value given that it ignores the information provided by the direction of the effect.

## 2. METHODS

The Receiver Operator Characteristic (ROC) curve of a test with a continuous outcome provides a link between the true positive rate (TPR), the false positive rate (FPR), and the likelihood ratio (LR) of a particular test result (Choi 1998). If the ROC curve is convex (monotonically increasing) and its slope is monotonically decreasing, then the following relationship exists:

$$LR = \frac{TPR - TPR^2}{FPR - FPR^2} \qquad (1)$$

where LR is the ratio of the likelihood that the condition being tested for is present versus the likelihood that it is absent. (See appendix A for justification of equation 1). Congruently, a trial can be approached in a similar manner with the data representing the result of a test, and the null hypothesis representing the condition being tested for (Cali and Longobardi 2015).

## *2.1 Likelihood Ratio of a One-sided P-value*

We consider a trial measuring a continuous variable ($\theta$) where the sampling distribution is symmetric (e.g. normal). *Pre-data,* an effect size of practical interest ($\Delta$) is identified and used to define dividing hypotheses, $H_1: \theta < \Delta$ and $H_2: \theta > \Delta$. In medical science for instance, $\Delta$ is usually defined as the minimum clinically important difference and is used in sample size calculations. It is not strictly necessary to specify whether $H_1$ or $H_2$ is the favored hypothesis pre-data (i.e. a directionality of effect), although doing so makes the bijection clearer: an effect in the favored direction produces an LR>1 and an effect in the opposite direction generates an LR<1. The trial is performed and an observed effect size ($\theta_{obs}$) is obtained. Using a ROC curve, any test value can be defined as a cut-off point between what is regarded as a positive or negative result, each cut-off having its own TPR and FPR values. Accordingly, *post-data* we declare a test threshold equal to $\theta_{obs}$, such that any effect size greater than or equal to $\theta_{obs}$ will be regarded as a "positive" test, that is, positive for the true effect size to be greater than $\Delta$. Note that this test is framed around the maximum likelihood estimate (MLE) of the true effect size, that is, the observed effect size. It is the optimal test threshold in that, a) any higher threshold would lead us to regard the test as 'negative' which is clearly wrong given the direction of the effect is obvious, and b) any lower threshold would fail to use the full magnitude of the observed effect size. A one-sided P-value is calculated against the dividing hypothesis ($\theta = \Delta$), the direction of which determines whether the data argue for the competing hypotheses $H_1$ or $H_2$. Under these circumstances, the one-sided P-value equals the FPR, being the probability of obtaining such a result if the true effect is in the opposite direction. To estimate the TPR, given that the observed effect size is always the maximum likelihood estimate, and that from equation (1) the likelihood ratio is maximised when TPR = 0.5, it follows that the likelihood ratio associated with a one-sided P-value is: $LR = \frac{0.25}{P-P^2}$, a likelihood ratio which we shall refer to as the maximum likelihood estimate likelihood ratio (MLE-LR), given that it is calculated from the TPR and FPR of the MLE-framed test. This is the ratio of the likelihood that the true effect size *equals* Δ against the likelihood that the true effect size is greater than Δ. The calculation maximises the LR and has parallels with maximisation of the Bayes factor as advanced by I.J. Good (1985, chapter 3). Generalising the likelihood ratio to the composite case, if the parameter space is divided at Δ then the supremum on one side is the likelihood of Δ, and hence the MLE-LR is the likelihood ratio that the true effect size lies on one side of the partition versus the other (Zhang 2009, Bickel 2012). That is, the LR against the simple hypothesis θ=Δ is the same as that against the composite (dividing) hypothesis θ<Δ.

It could be argued that while LR and hence TPR is maximal at the observed effect size, that maximum value might be less than 0.5. To address this, once again consider that any point (q) on the ROC curve can be used as a cut-off point for a test to determine whether we decide that the true effect size is in the same direction as the observed effect size, that is we declare the test is "positive" if $q_{obs} \geq q$. As above, if post-data we set q equal to the observed effect size then the test is always "positive" and the FPR of the test is the 1-sided P-value. Assuming the true effect size equals the observed effect size (the maximum likelihood estimate) and using this 1-sided P-value as the value for α in a power calculation then power=0.5=TPR. Here we are calculating the power (equivalent to TPR) using the observed effect size as the true effect size and the observed P-value as a, a procedure which always results in power=TPR=0.5. This phenomenon has previously been noted in discussions of observed (or post-hoc) power (Hoenig and Heisey 2001). Setting the alpha level equal to the P-value at first glance seems counter-intuitive but consider that the LR>1 must always be true given that it is calculated in the direction of the observed effect size. A power calculation always returns a value for power greater than or equal to the nominated alpha-level and therefore setting the alpha-level equal to the P-value ensures that TPR≥FPR and therefore LR≥1.

*2.2 Alternative Methods to Calculate the Likelihood Ratio (or Bayes Factor)*

Using the normal probability distribution function, the likelihood ratio against a point null has been calculated as described by Goodman (1999), after A.W.F Edwards: $LR = \exp\left(\delta^2/2\right)$, where δ is the observed effect size divided by its standard error. This is the likelihood ratio of $H_0: \theta = \Delta$ versus $H_A: \theta = \theta_{obs}$, which once again using the generalised ratio is equivalent to the likelihood ratio that the true effect size is greater than versus less than Δ. From Figure 1 it can be seen that this is less than the MLE-LR for all P-values. This is because the Goodman-Edwards likelihood ratio is not a bijection of δ given that $\delta^2 = (-\delta)^2$ and hence ignores the information contained in the knowledge of the direction of the effect, leading to an under-estimation of the likelihood ratio (Held and Ott 2018). In our simple normal-mean model, a trial of size N produces three pieces of information: the absolute effect size, the uncertainty of its estimation (proportional to $[\sqrt{N}]^{-1}$), and the direction of the effect. The Goodman-Edwards LR only uses the first two of these and thus is the likelihood equivalent of the 2-sided P-value.

We can view a two-sided P-value as being composed of two one-sided tests, one against $H_0: \theta < \Delta$ and the other against $H_0: \theta > \Delta$, the resultant 2-sided P-value being twice the value of the lesser of these two one-sided P-values. This makes it clear that a two-sided P-value is direction-agnostic. This formulation has obvious parallels with the TOST (two one-sided tests) procedure for testing equivalence – in a sense the two-sided P-value is a test for non-equivalence. For instance, if we consider a trial with a normal sampling distribution that returns a standardised effect size $\delta=1.96$, the two-sided P-value (the test of existence) is P=0.05, and the corresponding Goodman-Edwards likelihood ratio is LR=6.8. The one-sided P-value (the test of direction and magnitude) on the other hand is P=0.025 and the corresponding MLE likelihood ratio is LR=10.25. Finally, one may ask why it is not possible to calculate a MLE likelihood using a two-sided P-value. The problem is that such a P-value is composed of two one-sided tests, and it is unclear how the FPR and TPR from these two tests might be combined. Hence a one-sided P-value is compatible with likelihood-ratio based inference, but a two-sided P-value is not, as has been noted before (DeGroot 1973, Autzen 2016).

From the Bayesian perspective, going back historically as far as Gosset, the 1-sided P-value of itself has been sometimes regarded by Bayesians as approximately the posterior probability that the true effect is in the opposite direction to that observed, assuming an improper uniform prior (Greenland and Poole 2013), and if this is the case then LR=P/(1-P) (Marsman and Wagenmakers 2017). Alternatively, a second Bayesian method is the "e p log(p)" calibration of Selke, Bayarri and Berger (2001) which uses an arbitrary beta distribution as the prior. A comparison between these Bayesian approaches and the MLE-LR is shown in Figure 1.

Figure 1. The likelihood ratio/minimum Bayes factor associated with a 1-sided P-value (or its associated Z-statistic) from a normal distribution is shown for four different methods of calculation.

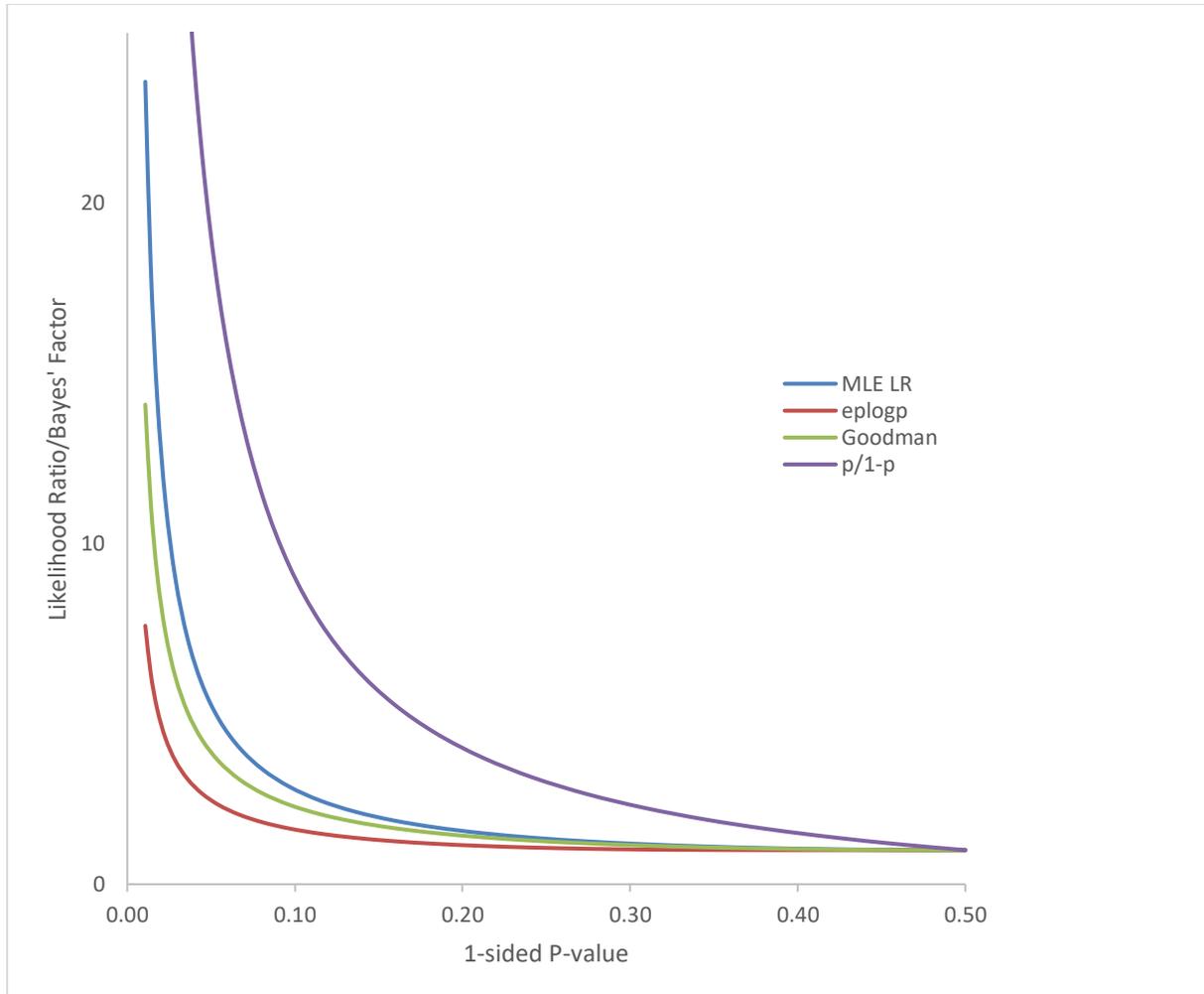

## 3. Cognitive Inference from a Likelihood Ratio

Defining a one-sided P-value as the (conditional) probability of obtaining the observed effect size if the true effect was in the opposite direction is, like the corresponding definition of a two-sided P-value, unhelpful and confusing to many (Goodman 2008, Chun Wah et al 2018, Reaburn 2017). A likelihood ratio on the other hand seems like a naturally comprehensible measure of the weight of evidence for or against a hypothesis (Good 1992). In addition, we can apply a form of Bayes theorem: $posterior\ odds = prior\ odds \times LR$, and estimate the posterior probability that the result represents a sign error (that is, an estimated effect in the opposite direction to the true effect). Of course, each individual end-user is free to assign their own prior odds which might be problematic, but it is somewhat simpler than assigning a full Bayesian prior probability distribution. Furthermore, calculating a one-sided P-value against an effect size of practical interest (Δ) rather than θ=0 gives a MLE-LR that represents a clear summary of the strength of evidence for or against the true value of θ being of practical significance. In addition, it is worth considering that if we assign prior odds of 1:1 (known as equipoise) then the posterior odds of the true effect being in the opposite direction to that observed are equal to the LR, a useful correspondence given that many non-statisticians have a good informal understanding of betting odds. This observation is not novel having been noted as far back as 1878 by C.S. Pierce but has perhaps been forgotten (Banks 1996).

As an illustrative example of the various methods of calculating an LR or minimum Bayes factor, we consider a poll that samples the voting intentions for an electoral contest between two candidates. The poll results favour one of the candidates with a proportion of 50% plus one standard error, and we ask what is the probability that this candidate will subsequently receive more than 50% of the vote assuming that these voting intentions are followed (Table 1).

Table 1. Comparison of four different likelihood ratios (LR) or Bayes factors (BF) for a dichotomous outcome calculated from a standardised effect size of δ=1, and their corresponding posterior probabilities in the absence of any prior knowledge.

| Method | Formula | Prior | LR/BF | Post. Probability |
|---|---|---|---|---|
| **Marsman et al (2017)** | P/(1-P) | Uniform | 5.25 | 84% |
| **MLE-LR** | $0.25/(P-P^2)$ | 1:1 odds | 1.86 | 65% |
| **Goodman (1999)** | $e^{(Z^2)/2}$ | 1:1 odds | 1.65 | 62% |
| **Selke et al (2001)** | -e p log(p) | Beta prior | 1.25 | 55% |

The probability distribution is binomial but as the proportion is close to 50% a normal approximation is appropriate. Intuitively, the posterior probability of about 2/3 seems appropriate whereas many would argue that 84% is too high (Gelman 2013), and 55% too low. As stated above, the Goodman (1999) likelihood ratio does not use all the information available and hence we can assume is too conservative.

## 4. DISCUSSION

A number of criticisms have been directed at P-values, particularly with the recent campaign to "abandon statistical significance" (Benjamin 2018). In this section we briefly argue that the majority of these criticisms do not apply to one-sided P-values, particularly when considered as bijections of a likelihood ratio. A detailed analysis and rebuttal, however, are beyond the scope of this article. Critical discussion of P-values can be easily found, for instance, in McShane et al (2019), Krueger and Heck (2017), and Szucs and Ioannidis (2017).

- The point null hypothesis is almost always implausible – this does not apply to a one-sided P-value against a dividing null hypothesis which instead implicitly assumes that the true effect size is non-zero.

- The incidence of sign errors is large (and hidden) – the probability of a sign error (FPR) is integrated into the calculation of the MLE-LR and is explicit.

- Multiple comparisons invalidate the P-value – if the likelihood principle holds then the one-sided P-value expressed as a likelihood ratio should not require adjustment for multiple comparisons nor multiple looks at the data.

- The P-value is a poor measure of the strength of evidence for or against a hypothesis – a likelihood ratio is often regarded as the ideal measure of strength of evidence (Edwards 2001, Royall 1997).

- The P-value invites confusion between statistical significance and practical importance – selecting an appropriate value of $\theta$ for the dividing null hypothesis $H_0<\Delta$ or $H_0>\Delta$ will lead to the one-sided P-value and its associated MLE-LR directly addressing the likelihood of an effect size of practical importance.

- The P-value is difficult to understand – a LR can be combined with a prior probability to produce an explicit and clear posterior probability that its associated hypothesis is true.

- Point and dividing hypotheses are forms of interval hypotheses and P-values are incoherent when calculated for an interval hypothesis nested within a second interval hypothesis (Schervish 1996) – dividing hypotheses cannot be nested inside each other given that they extend to $+\infty$ or $-\infty$. Furthermore, Lavine and Schervish (1999) demonstrate that likelihood ratios are coherent whereas Bayes factors are not.

- One-sided P-values as a test of direction are invalid if the parameter value is exactly zero, on the grounds that in this case the test is between two directional models that are both wrong (Marsman and Wagenmakers 2017) – there is no substantial difference between calculating the 1-sided P-value against a $H_0: q > 0$ versus $H_0: q \geq 0$.

Goodman (2008) advocates use of Bayes factors instead of two-sided P-values and includes a table (page 139) summarizing desirable properties of an evidential measure. He claims all of these properties are possessed by the Bayes factor, and none by the two-sided P-value. The one-sided P-value, however, possesses all these desirable evidential qualities when considered as a likelihood ratio, a fact that is hardly surprising given the close relationship between Bayes factors and likelihood ratios. These qualities include providing information about the effect size, using only observed data, including an explicit alternative hypothesis, insensitivity to stopping rules, and being a measure that increases (rather than decreases) as the strength of evidence increases.

In summary, a common and simple type of research is the comparison of two different interventions where the aims are to determine a) which intervention is better, and b) what is the strength of evidence supporting this determination. Null hypothesis significance testing using a 2-sided P-value seems ill-suited to this task whereas, viewed as a bijection of a likelihood ratio, the one-sided P-value can be incorporated into the likelihood evidential paradigm (see Royall 1997). We would argue that this paradigm is simpler to comprehend and more coherent than that of null-hypothesis significance testing using a two-sided P-value. Resurrection of the one-sided P-value mitigates many of the problems that have led to the call to abandon the use of P-values.

Appendix: Justification of Equation 1.

Consider a trial where a sample is taken, the mean ($\mu$) of which is normally distributed around the true population mean, and a dividing hypothesis such as $H_0$: $\mu<0$ is specified. It is possible to specify a cut-off value for $\mu$ that determines whether to reject or accept $H_0$. Each cut-off value has an associated TPR and FPR and from these a ROC curve can be constructed. Under these conditions, the ROC curve will be convex from above (or proper), monotonically increasing and with a monotonically decreasing slope. Two secants to the curve can be drawn from the specified cut-off value (A) to the origin (0,0, bottom left) and to the terminus (1,1, top right) (Figure 2).

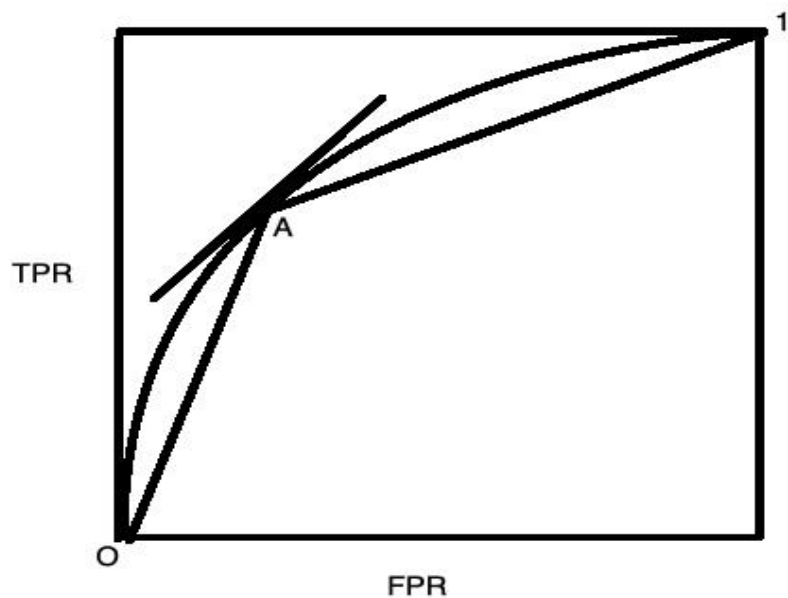

Figure 2. A proper ROC curve. The slope of secant 0-A equals the positive likelihood ratio. The slope of secant A-1 equals the negative likelihood ratio. The slope of the tangent to the curve at point A equals the likelihood ratio at that point.

The slope of each these secants equals the average slope of their corresponding curve segments, and also the positive and negative likelihood ratios respectively, associated with the specified cut-off value. The average slope of the ROC curve as a whole always equals one, as shown by a diagonal (0,0 to 1,1) and hence the average slope of the two secants combined is also equal to one.

If we convert the cut-off point A to a very small interval (A ε ±) bridged by a third secant, the slope of the two original secants will remain *approximately* the same, and, as the average slope of the overall ROC curve is unchanged, the slope of the third secant must be the product (by the chain rule) of the slopes of these two original secants. Taking the limit as ε approaches zero, the approximation disappears, and the slope of the now infinitesimal secant equals the tangent to the curve at point A:

$$\lim_{\varepsilon \to 0}(LR) = LR^+ \times LR^- = \frac{TPR}{1-FPR} \times \frac{1-TPR}{FPR} = \frac{TPR - TPR^2}{FPR - FPR^2}$$